\documentclass[aps,prl,groupedaddress,twocolumn,showpacs,preprintnumbers,amsmath,amssymb]{revtex4}

\usepackage{graphicx}
\usepackage{dcolumn}
\usepackage{bm}

\begin{document}

\preprint{APS/123-QED}

\title{Reverse Micelles Enable Strong Electrostatic Interactions of Colloidal Particles in Nonpolar Solvents}

\author{M. F. Hsu}
 \altaffiliation[Now at ]{General Electric Global Research,
 Niskayuna, NY}
\author{E. R. Dufresne}
 \altaffiliation[Corresponding author.  Now at ]{Dept. of Mechanical Engineering,
 Yale University, New Haven, CT.}
 \email{eric.dufresne@yale.edu}
\author{D. A. Weitz}
\affiliation{Dept. of Physics and DEAS, Harvard University,
Cambridge, MA 02138, USA}

\date{\today}

\begin{abstract}

While the important role of electrostatic interactions in aqueous
colloidal suspensions is widely known and reasonably
well-understood, the relevance of charge in nonpolar suspensions
remains mysterious.  We demonstrate that particles can have
surprisingly strong electrostatic interactions in low dielectric
constant environments when ions are solubilized by reverse
micelles.  A simple thermodynamic model, relating the structure of
the micelles to the equilibrium ionic strength, is in good
agreement with both conductivity and interaction measurements.
Since dissociated ions are solubilized by reverse micelles, the
entropic incentive to charge a particle surface is qualitatively
changed, and surface entropy plays an important role.

\end{abstract}

\maketitle


Charge is a powerful tool for manipulating the properties of
complex fluids.  In aqueous colloids, charges readily dissociate
and play an essential role in determining the structure of
suspensions and their response to electrical, mechanical and
optical stimuli. Perhaps most importantly, charge stabilizes
aqueous colloidal particles against flocculation.  By contrast,
charge is commonly thought to play no role in nonpolar
environments in thermodynamic equilibrium. However, a close
examination of existing technologies suggests a richer story. Most
stunningly, by assigning opposite charges to absorbing and
reflecting particles, nonpolar suspensions have been engineered to
act as electrophoretic ink \cite{comiskey.1998}. Furthermore,
charge control agents have long been added to petroleum products
to reduce the risk of explosions during processing and transport
\cite{klinkenberg.1958}. There is even some evidence that charge
contributes to the stability of soot in diesel engine oil
\cite{pugh.1983}. In these cases, reverse micelles play a crucial
role by reducing the electrostatic penalty of ionization and
thereby stabilizing ions in equilibrium \cite{morrison.1993}. In
all cases, it is absolutely essential that ions are made soluble
in this low dielectric constant environment. Recently, force
measurements have provided the first direct evidence of
electrostatic repulsions between charged surfaces in a nonpolar
solvent \cite{briscoe.2002,mcnamee.2004}.
 Despite the growing acknowledgement of the importance of charge in
nonpolar systems, a useful quantitative framework to describe the
origin and interactions of charge, comparable to the theory of
Derjaguin-Landau-Verwey and Overbeek (DLVO) for aqueous systems
\cite{derjaguin.1941,verwey.1948}, does not exist. Such an
understanding is essential to fully exploit the technological
capabilities of charge in nonpolar solvents and to fully
comprehend the behavior of these fascinating systems

In this Letter, we present a physical picture of the origins of
charge and the interactions of colloidal particles in nonpolar
solvents with reverse micelles.  Our experiments show that
particles charge and interact strongly with screened-Coulomb
interactions. Ions reside within the polar cores of reverse
micelles.  Reverse micelles not only stabilize charge dissociated
from particle surfaces, but also charge spontaneously to provide
additional screening of electrostatic interactions.  The
confinement of charge within reverse micelles qualitatively
changes the entropic incentive to charging: the configurational
entropy of the particle surface charge, typically negligible in
aqueous systems, plays a critical role.

We explore the interactions of colloidal PMMA particles suspended
in a mixture of oil and surfactant. The particles have a radius of
780 nm and are sterically-stabilized by a layer of
poly(12-hydoxy-steric-acid)
 (PHSA) hairs grafted to their surface.   The particles are suspended in a
mixture of dodecane and an ionic
surfactant,di-2-ethylhexylsulfosuccinate (AOT). Above its critical
micellar concentration, AOT forms reverse micelles whose size is
very nearly independent of concentration,
\cite{mathews.1953,mukherjee.1993}. These micelles are very
hygroscopic. To minimize the effects of moisture, we mix pure
dodecane (99\%) and AOT (98\%) in a dry glovebox. All measurements
are either performed in the glovebox or with sample cells that
have been sealed in the glovebox.

In pure dodecane, our PMMA particles reversibly aggregate, as
shown in Fig. 1(a). \begin{figure}
\includegraphics[scale=0.9]{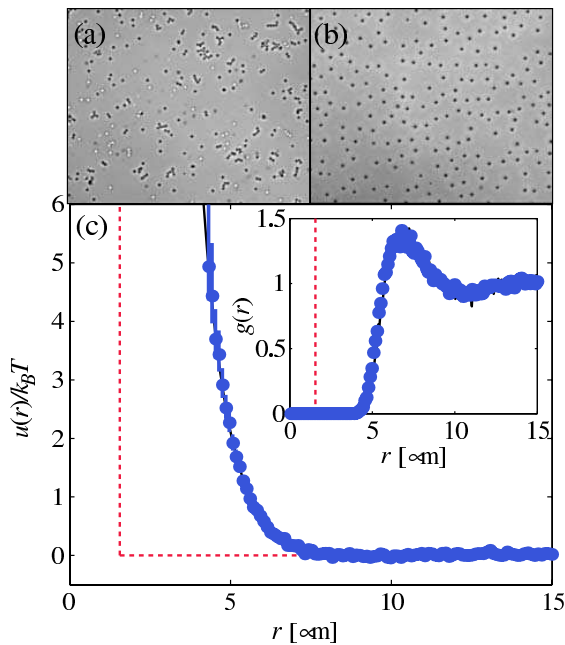}
\caption{ Charge-stabilization of a nonpolar suspension. (a)
Optical micrographs of PMMA particles in pure dodecane and (b)
with 12mM AOT, field of view $135 \times 108$ $\mu$m$^2$. (c)
Inferred particle pair potential, symbols, and fit to
screened-Coulomb interaction, line. For comparison, a hard sphere
interaction is plotted as a dashed line. Inset. Measured and
simulated equilibrium pair correlation functions, symbols and
line, respectively. A dashed line indicates contact.}
\end{figure} However, when AOT is added at concentrations
above its critical micellar concentration, the particles disperse.
Moreover, particles in suspensions with 12 mM AOT rarely approach
within three particle diameters in each other, as shown in Fig.
1(b).  We quantify this surprisingly strong repulsion by
collecting a series of statistically independent images of a
freely-diffusing monolayer of particles confined between two
microscope slides. The centers of each particle are then precisely
located \cite{crocker.1996a} to calculate the ensemble-averaged
two-point density correlation function, $g(r)$, using a total of
about 40,000 particles. As shown in the inset of Fig. 1(c),
particles in a 12 mM suspension never come within four microns of
each other. At larger separations, $g(r)$ rises slowly to a peak
near 7 $\mu$m and decays to unity at larger separations. In order
to minimize the effect of the confining surfaces, the surfaces of
the sample cell are prepared to mimic the surface of the particles
and interactions are measured in the thickest regions of the cell
that contain a well-confined monolayer \footnote{Out-of-plane
motion leads to systematic errors in the particle-particle
separation of less than three percent. Systematic variations in
the forces between the particles due to interactions with the
walls will be described elsewhere. }

The observed pair-correlation function is consistent with the
repulsive component of the DLVO theory of colloidal interactions.
We extract pair potentials by inverting $g(r)$ with liquid
structure theory \cite{behrens.2001}. Using the resulting
potentials, 2-D Monte Carlo simulations of our experiments
generate pair-correlation functions that are in good agreement
with those measured. A comparison of measured and simulated
$g(r)$'s for a sample with 12 mM AOT is shown in the inset of Fig.
1(c), where data and simulation are indistinguishable. The
observed potential, shown in Fig. 1(c), is in excellent agreement
with the screened-Coulomb form,
\begin{equation} \frac{u(r)}{k_BT}=\left(\frac{e\zeta}{k_BT}\right)^2
\frac{a^2}{\lambda_B}\frac{e^{-\kappa(r-2a)}}{r},\end{equation}
where $\zeta$ is the apparent potential at the surface of the
particle.  The Debye length, $\kappa^{-1}=1/\sqrt{4 \pi \lambda_B
n_{ion}}$, characterizes the range of electrostatic interactions,
where $n_{ion}$ is the number density of ions. The Bjerrum length,
$\lambda_B = e^2/4\pi \epsilon \epsilon_o k_BT$, is the separation
of two elementary charges when their electrostatic interaction
energy is equal to $k_BT$. A least-squares fit to our data yields
$e|\zeta|/k_BT=4.4\pm0.3$ and $\kappa=(1.07\pm0.05)$
$\mu$m$^{-1}$. Defying conventional wisdom, this zeta potential is
as high as those found
 in highly-charged aqueous systems \cite{behrens.2000,crocker.1994}.  However, the
  effective charge, \begin{equation}\label{zzeta}Z=\zeta (a/\lambda_B) (1 + \kappa a),\end{equation}
 is $225\pm15$,  two orders of magnitude below
 that measured in similarly sized charged particles in water.
 This charge on the surface of particles must be balanced by counterions
 in solution to ensure charge neutrality.  Using
 our measured value of $Z$ and the number density of particles, we
 calculate the density of counterions to be about $0.5$ $\mu$m$^{-3}$.
 However, this ion density is too low to account for the
 observed value of $\kappa$, which implies a significantly larger ion density of $3.3 \pm 0.3$ $\mu$m$^{-3}$.
 Therefore, particle surfaces must not be the only source of stable ions in solution.

Indeed, stable ions exist in solutions of AOT in dodecane without
particles. While these ions are too small to be imaged optically,
we can detect their presence through their contribution to
solution conductivity. Assuming that charge is carried by singly
charged micelles,
 then the conductivity \begin{equation} \label{eq:cond} \sigma=\frac{e^2n_{ion}}{6\pi \eta a_h},\end{equation}
 where $a_h$ is the hydrodynamic radius of the micelles and $\eta$ is the solvent viscosity.
  Our 12 mM AOT solutions have a conductivity of
 17 pS/cm.  Using the known hydrodynamic radius of AOT reverse
 micelles in dodecane, we find an ion density of $2.7 \pm 0.4$ $\mu$m$^{-3}$,
 consistent with $\kappa = 0.98 \pm .05$ $\mu$m$^{-1}$, in
 good agreement with our interaction measurements.
 Therefore, ambient charged micelles dominate the ionic strength
 of the system.  However, by comparing the number density of ions to the
 number density of micelles, calculated from AOT concentration and micellar
 aggregation number, we find that only one micelle in $10^5$
 develops a charge.

To elucidate the origin of this excess charge, we measure the
scaling of solution
 conductivity with AOT concentration.  Varying
 the concentration of AOT from 0.8 to 200 mM, the conductivity
 increases linearly from 1 to 500 pS/cm, as shown in
 Fig. 2. \begin{figure}
\includegraphics{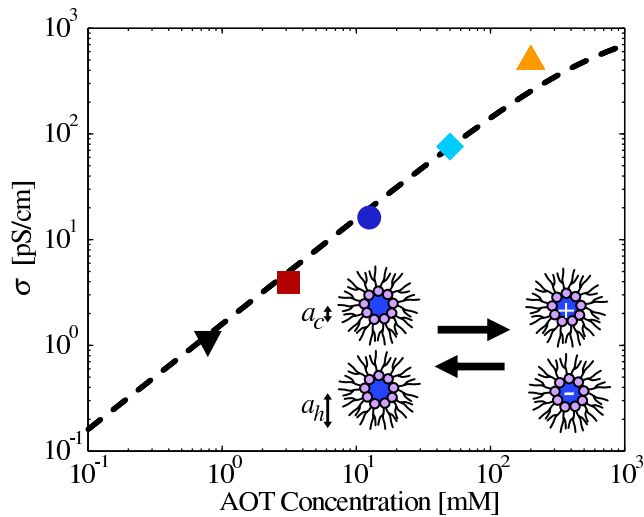}
\caption{ \label{cond} Conductivity of AOT/dodecane solutions
without particles. Symbols indicate measurements, and dashed curve
indicates fit to Eq. \ref{eq:cond} with empirical [AOT]-dependent
solution viscosity. \emph{inset}  Schematic diagram of two-body
process that leads to the creation of charge in the bulk of the
solution.}
\end{figure}  Therefore, the ratio of ions to micelles,  $\chi=n_{ion}/n_{micelle}$
 is indepedent of the concentration of AOT, and  $\chi=1.2 \times 10^{-5}$.
  Only a tiny fraction of micelles ionize to form a
  charge carrier.  While weak electrolytes, of
 the form $AB \rightleftharpoons A^{+}+B^{-}$, can exhibit similar
 ionization fractions, $\chi$ would scale like $[AB]^{-\frac{1}{2}}$.
 Instead, a charging mechanism where neutral micelles
 reversibly exchange charge though a collision, $A+A
 \rightleftharpoons A^++A^-$, leads, through the law of mass
 action, to a ionization fraction independent of the concentration
 of $A$ \cite{eicke.1989, morrison.1993}, as demanded by our data.  The fraction of charged micelles is determined by the
 difference in free energy between the charged and uncharged states, $\chi=2 \exp(-\Delta F/2k_BT)$.
 If the free-energy cost of ionizing a single isolated micelle is $g_M$, then $\chi=2 \exp(- g_M/k_BT)$.
 Comparing this basic thermodynamic result to our measured value of
 $\chi$, we find that $g_M=12$ $k_BT$.

 The measured micelle ionization
 energy has a simple electrostatic interpretation.  A singly-charged
 conducting sphere of radius $a_c$ surrounded by a dielectric
 medium has a electrostatic energy $\lambda_B/2a_c$ \cite{parsegian.1969}.
 Assuming that electrostatic contributions dominate the
 ionization free-energy and the core of the micelle is effectively a conductor,
 then our measured value of $g_M$ implies a core radius,
 $a_c=1.2$ nm.  This is in good agreement with neutron scattering
 measurements \cite{kotlarchyk.1985}.  Thus, we conclude that the
 equilibrium ionic strength of a solution of reverse micelles is determined by a competition between entropy and
 electrostatics.  The relative size of the Bjerrum length and the
 micellar core radius determine the charge fraction.  When $\lambda_B\gg a_c$ a small fraction of
 micelles charge.  When $\lambda_B \approx a_c$, nearly all of the
 micelles can charge and individual micelles can become multiply-charged.

\begin{figure}
\includegraphics{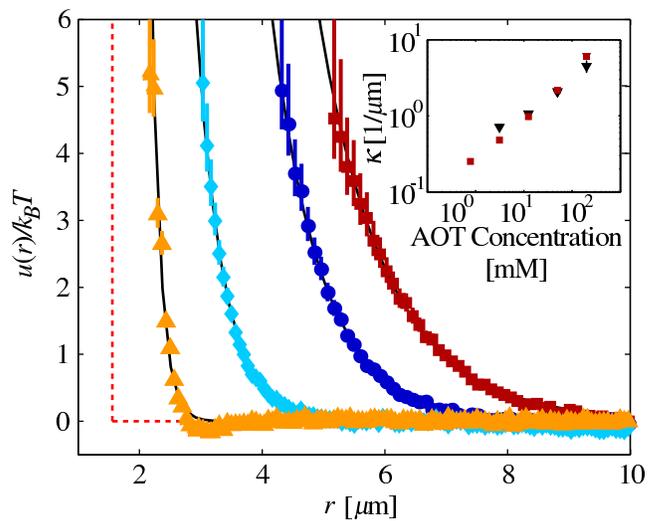}
\caption{ \label{aot} Variation of interactions with [AOT].
Inferred interactions for solutions of 3, 12.5, 50 and 200 mM AOT,
 squares,  circles,  diamonds and triangles,
respectively.  A hard sphere potential is shown as a dashed line
for comparison.  Screened-Coulomb fits to each potential are
plotted as solid lines. \emph{inset} Comparison of $\kappa$ as
inferred from conductivity and equilibrium structure, crimson
squares and black diamonds, respectively.}
\end{figure}

 Ionic strength has a dramatic effect on the interactions between particles.
 We find that the range of the interactions is greatly reduced
 as the concentration of AOT increases from 3 to 200 mM, as shown in Fig. 3.
 At the lowest concentration of AOT, particles feel a repulsion greater
 than $k_BT$ at center-to-center separations less than 7 $\mu$m
 or $4.5$ particle diameters. At the highest concentration,
 particles can approach each other within 2.5 $\mu$m.
 These data are all fit by the DLVO
 form with a constant apparent surface potential $e|\zeta|/k_BT=4.1\pm0.1$
 and $\kappa^{-1}$ ranging from $0.2$ to $1.4$  $\mu$m.  The ionic
 strengths
 inferred from fits to interparticle potentials are in good agreement with the ionic
  strengths inferred from conductivities, as shown in the inset to
  Fig. \ref{aot}.   However, the discrepancy between these two measures widens at the lowest
 concentration of AOT, reflecting the contributions of
counterions to screening. Interestingly, we consistently
 observe a very small attraction in 200 mM samples.  We
 believe that this attraction reflects the underlying structure of
 the reverse micelle solution, which displays a pronounced increase
 in viscosity at this concentration.

Our conductivity and interaction measurements show that
thermodynamics is an excellent tool for predicting the
concentration of ions in solution. Thermodynamics can also
describe the charging of colloidal particles.  In order for a
particle to become charged in equilibrium, the energy cost of
creating and separating surface charges and counterions must be
overcome by entropy. This energy cost has three components.  As we
have already shown, the counterion has a solvation energy,
$g_M=12k_BT$. Similarly, an ion on the particle's surface has a
solvation energy, $g_S$. Once the ion and counterion have been
created, they must be separated at an energy cost of
$e\left|\zeta\right|$, where $\zeta$ is the electrostatic
potential at the surface of the particle. These costs are overcome
by gains in the configurational entropy of micelles in the bulk
and the configurational entropy of charges on the particle
surface. When the particle charge increases by one, a neutral
micelle in the bulk is converted to a charged micelle. Thus, the
change in the free energy is $k_BT \ln [n_{ion}/2
n_{micelle}]=k_BT \ln [\chi/2]$, where we have assumed that only a
tiny fraction of micelles are charged, and that the entropies of
the charged and neutral ionic species have the same form as an
ideal gas. If a particle has $N$ chargeable sites and a charge of
$Z$ with  $Z\ll N$, then the surface entropy's contribution to the
chemical potential is $k_BT \ln \left[Z/N\right]$.
 Thus, the total chemical potential is
\begin{equation} \frac{\mu}{k_BT} = g_S + g_M +
\frac{e|\zeta|}{k_BT}+\ln\left[\frac{Z}{N}\right]+\ln\left[\frac{\chi}{2
}\right].
\end{equation}   In
equilibrium, $\mu=0$.  In the limit where excess ions dominate
over counterions, $\chi=2 \exp(-g_M)$, thus
\begin{equation} e|\zeta| \approx -g_S-\ln[Z/N].
\end{equation}  Remarkably, the bulk entropic contribution to the chemical potential
 exactly cancels the solvation energy of the counterion.  This
 surprising cancellation occurs because $\chi$ is pinned by the solvation energy $g_M$,
 but
 will breakdown when a significant fraction
 of the charged micelles contain counterions.
 Consequently, the surface potential depends primarily on the
competition between surface entropy and the solvation energy of an
ion on the surface.  However, since $Z$ depends on the ionic
strength, Eq.(\ref{zzeta}), there is some residual dependence on
$g_M$.  Since the left hand side is positive definite, meaningful
results are obtained only when $-\ln [Z/N] > g_S$. Therefore,
surface entropy's contribution to the chemical potential must
overcome the solvation energy of the surface charge.  This is in
stark contrast to thermodynamic models of charging in aqueous
systems, where the configurational entropy of the surface is
typically ignored \cite{alexander.1984}.  

Combining this relation with our structural and electrokinetic
measurements, we can estimate the solvation energy of a charge on
the particle surface, $g_S$. Assuming that charges on the surface
are constrained to reside within an adsorbed micelle, then packing
constraints limit the number of charges to $10^6$. Our interaction
measurements are consistent with $e \zeta /k_BT\approx 4$ and
$Z\approx 100-1000$, therefore $g_S \approx 3-5$ $k_BT$, about
one-quarter to one-half of the energy cost of charging an ion in
bulk.

We have found that the electrostatic interactions of colloidal
particles in a nonpolar solvent can be surprisingly strong and
tunable.   While the functional form of interparticle potentials
is identical the those of charged particles in aqueous solvents,
the thermodynamics of charging is qualitatively different.  We
expect reverse-micelle mediated colloidal interactions to play an
important role in the stability and structure of new colloidal
systems and to enable the exploration of fundamental physical
phenomena in previously unattainable regimes. Moreover, our
framework may help to illuminate the origin and consequences of
charging in technologically relevant nonpolar systems.

\begin{acknowledgments}
We thank Ian Morrison, Craig Herb, Phil Pincus and Paul Chaikin
for helpful discussions. We thank Infineum for the donation of
equipment and E*Ink for access to their light-scattering
facilities. We thank Sven Behrens and David Grier for sharing
their liquid structure code and Daniel Blair for help with the
Monte Carlo simulations. We thank Andrew Schofield for providing
the PMMA particles. We thank NASA (NAG 3-2284), NSF (DMR-0243715)
and the Harvard MRSEC (DMR-0213805) for funding.
\end{acknowledgments}


\end{document}